\documentclass[twocolumn,secnumarabic, amssymb, superscriptaddress, nobibnotes, aps, prl]{revtex4-2}

\usepackage{graphicx}
\usepackage{dcolumn}
\usepackage{bm}
\begin{document}

\title{Highly correlated optomechanical oscillations manifested by an anomalous stabilization}%
\author{Jinlian Zhang}%
\affiliation{Fujian Key Laboratory of Light Propagation and Transformation \& Institute of Systems Science, 
	College of Information Science and Engineering, Huaqiao University, Xiamen 361021, China}
\author{Miguel Orszag}%
\affiliation{Center for Quantum Optics and Quantum Information, Universidad Mayor, Camino La Pir\'{a}mide 5750, Huechuraba, Chile}	
\author{Min Xiao}
\affiliation{National Laboratory of Solid State Microstructures, College of Engineering and Applied Sciences, School of Physics, Nanjing University, Nanjing 210093, China}
\author{Xiaoshun Jiang}
\email[ ]{jxs@nju.edu.cn}
\affiliation{National Laboratory of Solid State Microstructures, College of Engineering and Applied Sciences, School of Physics, Nanjing University, Nanjing 210093, China}		
\author{Qing Lin}%
\email[ ]{qlin@hqu.edu.cn}
\affiliation{Fujian Key Laboratory of Light Propagation and Transformation \& Institute of Systems Science, 
	College of Information Science and Engineering, Huaqiao University, Xiamen 361021, China}
\author{Bing He}%
\email[ ]{bing.he@umayor.cl}
\affiliation{Center for Quantum Optics and Quantum Information, Universidad Mayor, Camino La Pir\'{a}mide 5750, Huechuraba, Chile}

\begin{abstract}
Driven by a sufficiently powerful pump laser, a cavity optomechanical system will stabilize in coupled oscillations of its cavity field and mechanical resonator. It was assumed that the oscillation will be continuously magnified upon enhancing the driving laser further. However, based on the nonlinear dynamics of the system, we find that the dynamical behaviors of the system are much more complex than this intuitive picture, especially when it is operated near the blue detuning point by the mechanical resonator's intrinsic frequency. There exists an anomalous stabilization: depending on its intrinsic damping rate and the pump power, the mechanical resonator will metastably stay on one orbit of oscillation after another until it completely stabilizes on the final orbit it can reach. These orbits are consistent with the locked ones with almost fixed oscillation amplitudes, which are realized after the pump power becomes still higher. The oscillatory cavity field is seen to adjust its sidebands following the mechanical frequency shift due to optical spring effect, so that it always drives the mechanical resonator to near those locked orbits once the pump power is over a threshold. In the regimes with such correlation between cavity field sidebands and mechanical oscillation, the system's dynamical attractors are confined on the locked orbits and chaotic motion is also excluded. 
\end{abstract}
\maketitle

Multiple oscillators coupled through nonlinear interaction potential exhibit nontrivial phenomena such as phase synchronization \cite{syn01,syn02} and internal resonance of energy transfer \cite{inter1, inter2}. When it comes to a pair of celestial bodies (two generalized oscillators), a complicated scenario can emerge: by its appropriate deformation and the adjustable gravitational interaction from the other, one of them will have its rotation and spin velocity locked to a fixed ratio (spin-orbit resonance) \cite{tidal}, explaining why the same side of the Moon always faces the Earth. Here, we illustrate that, through a unique mechanism realized by radiation pressure, a mechanical simple harmonic oscillator can be locked to a set of fixed orbits with its shifted frequency and a correspondingly adjusted light field spectrum.

\begin{figure}[t]
	\centering\includegraphics[width=\linewidth]{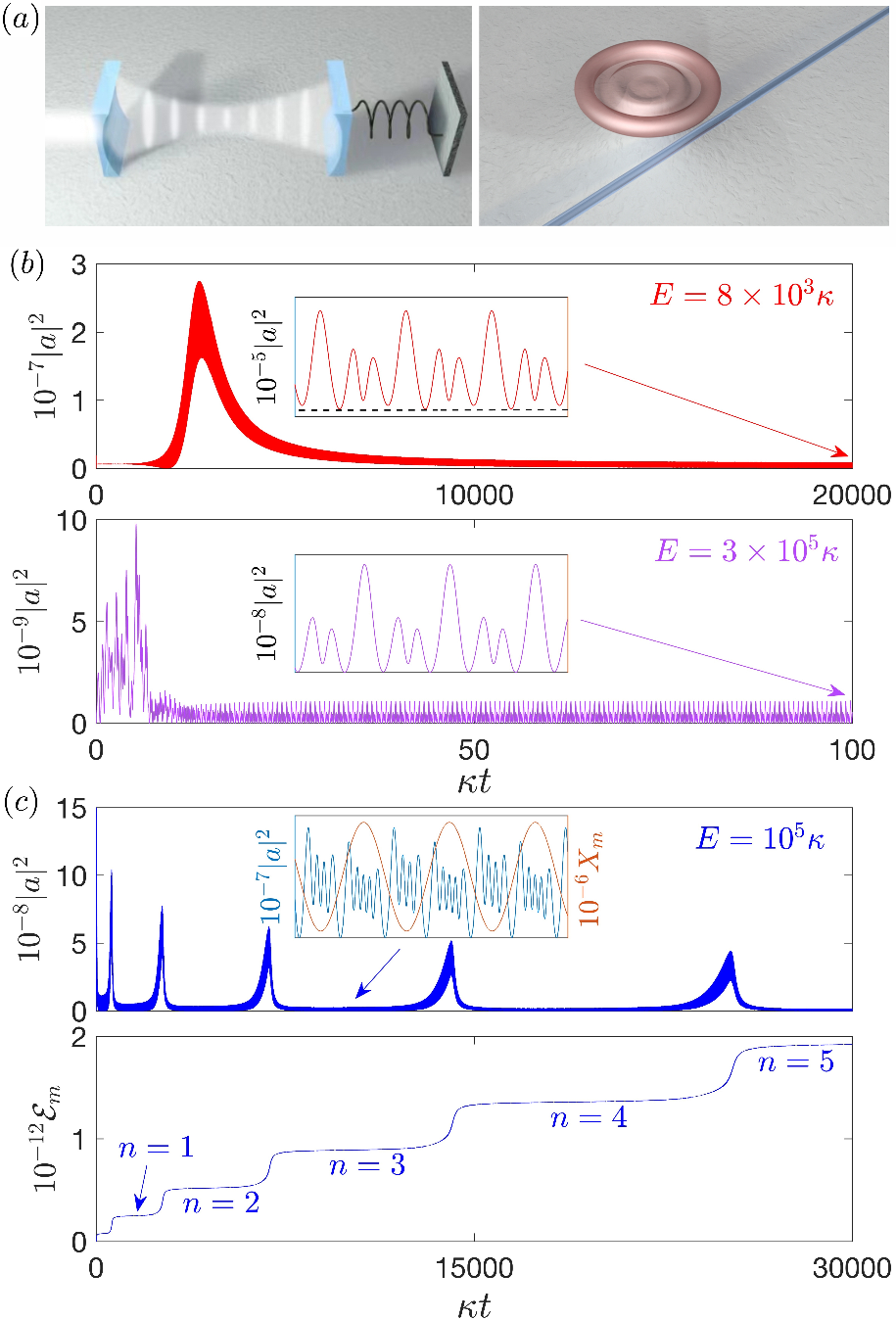}
	\caption{(a) Two exemplary types of COMS: a movable mirror of Fabry-Perot cavity and the breathing mode of a microcavity driven through an optical fiber. (b) The normal stabilization processes of the intracavity photon number outside the
	AS regime and exactly at the RPB ($\Delta=-\omega_m$). The dashed line in the inset of the upper panel highlights a feature that the field oscillation does not split into two distinct parts in each whole period. (c) An anomalous stabilization at the RPB. Due to the pulsed $|a(t)|^2$ in succession, the metastable cavity field and mechanical oscillation repeatedly change together with time, manifesting the mechanical energy steps $n=1,2,3$ and others. The lower (upper) limit for the AS phenomenon is around $E=9.5\times 10^3\kappa$ ($1.711399\times 10^5\kappa$). Together with the indicated drive amplitudes $E$, the system parameters are all scaled by the damping rate $\kappa$: $g_m=10^{-4}\kappa$, $\omega_m=10\kappa$, and $\gamma_m=0$.}  
	\label{fig1}
\end{figure}

\begin{figure}[b]
	\centering\includegraphics[width=0.99\linewidth]{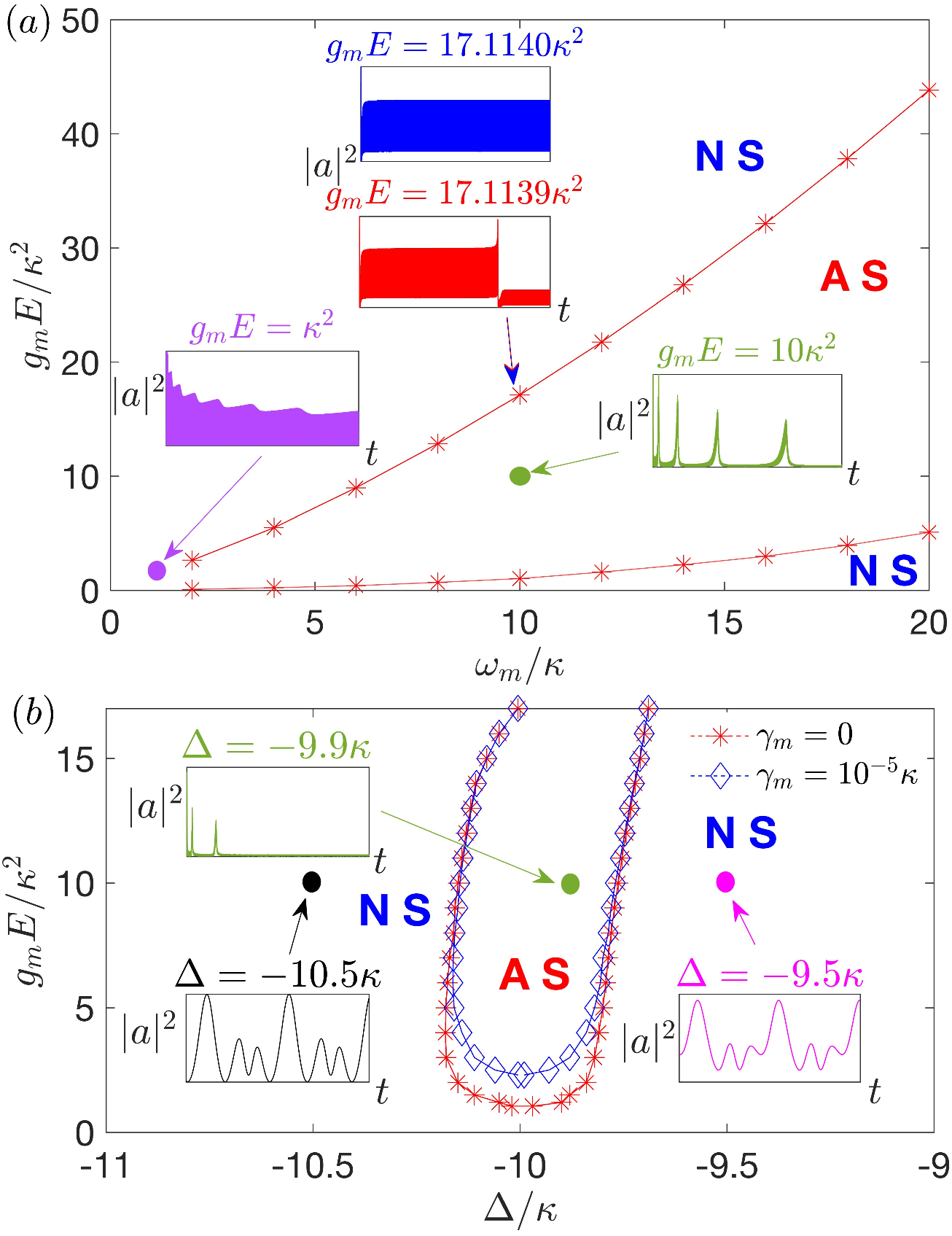}
	\caption{(a) The distribution of the AS phenomenon in the space ($\omega_m/\kappa$, $g_mE/\kappa^2$). 
		The green and purple insets based on the parameters in Fig. 1 show the cavity field evolution over the duration of $\kappa t\in[0,2\times 10^4]$ at the indicated points, and a transition across the upper boundary (the blue and red insets) is demonstrated within a duration $\kappa t\in[0,5\times 10^4]$. Here the system driven at the RPB is the one with $\gamma_m=0$. (b) The distribution of AS along the pump detuning. It is at the fixed $\omega_m/\kappa =10$ in (a), where another exemplary system with $\gamma_m=10^{-5}\kappa$ has the approximate lower (upper) boundary at $g_mE/\kappa^2=2.296$ $(17.1139)$, while its Hopf bifurcation is around $g_mE/\kappa^2=0.2$. The duration of the green inset is still between $\kappa t=0$ and $2\times 10^4$, and those of two others are slightly over two mechanical oscillation periods.}  
	\label{fig2}
\end{figure}

So far, the effects of radiation pressure have been widely studied with cavity optomechanical systems (COMS) \cite{rev1, rev2} exemplified in Fig. \ref{fig1}(a). Those effects are asymmetric with respect to the difference (detuning) of the driving laser frequency $\omega_l$ from the resonance frequency $\omega_c$ of the associated optical cavity. Scanning the pump frequency from the point red-detuned by the mechanical frequency $\omega_m$ of a COMS, where the system can stabilize to a static equilibrium so that optomechanical cooling can be implemented \cite{cool1,cool2,cool3, cool4,cool5,cool6,cool7,cool8}, to where it is blue-detuned by the same $\omega_m$, one would encounter optomechanical oscillations \cite{os1,vibration,os2,os3,os4,os5,os6,os7,os8,os9,os10,os11,os12,omcomb} on the way: the mechanical resonator will stabilize in oscillation, while the cavity field is modulated to have more frequency components as the sidebands. On the other hand, a Hopf bifurcation determined by the pump power exists at a fixed drive frequency, as the boundary between static equilibrium and optomechanical oscillation. Viewed from the linearized dynamics, a two-mode squeezing effect is enhanced at the exact blue-detuning by the mechanical frequency $\omega_m$ \cite{rev2}, which we call resonance point of blue-detuning (RPB), and it could be beneficial for generating optomechanical entanglement \cite{eta,etb,etc,etd,T-entangle}. A more powerful pump driving at a RPB was thus expected to always magnify the optomechanical oscillation further. However, based on the full nonlinear dynamics, the system driven at the RPB is found to stabilize in an anomalous way when the pump power is between one threshold higher than the Hopf bifurcation point and another transitional value towards a different regime of dynamics, resulting in a behavior contrary to the intuition that a stronger pump simply leads to more augmented optomechanical oscillation. 

For a realistic cavity, its displacement $x_m$ under radiation pressure is much less than its original size, giving rise to an interaction potential $V(X_m)=-\hbar g_m X_m|a|^2$ for the coupled cavity field and mechanical resonator, where $|a|^2$ is the intracavity photon number and $X_m=\sqrt{\frac{m\omega_m}{\hbar}} x_m$ ($m$ is the mechanical resonator's effective mass) is dimensionless. Then, the dynamical equations of the COMS are written in the reference frame rotating at the frequency $\omega_c$ as (see Appendix I)
\begin{eqnarray}
	\dot{a}&=&-\kappa a+ig_m X_m a+E e^{i\Delta t},\\
	\ddot{X}_m&=&-\gamma_m \dot{X}_m-\omega_m^2 X_m+g_m\omega_m |a|^2,
\end{eqnarray}
where $\Delta=\omega_c-\omega_l$ is the laser detuning and its pump power $P$ specifies the drive amplitude $E=\sqrt{2\kappa_eP/(\hbar \omega_l)}$. The mechanical damping rate $\gamma_m$ can be much smaller than the field decay rate $\kappa=\kappa_e+\kappa_i$ from both pump-cavity coupling and intrinsic loss. To obtain an exact picture of dynamics, we will stick to the numerical calculations with Eqs. (1) and (2), together with some analytical arguments for interpreting the results. 

\begin{figure*}[t]
	\centering\includegraphics[width=\linewidth]{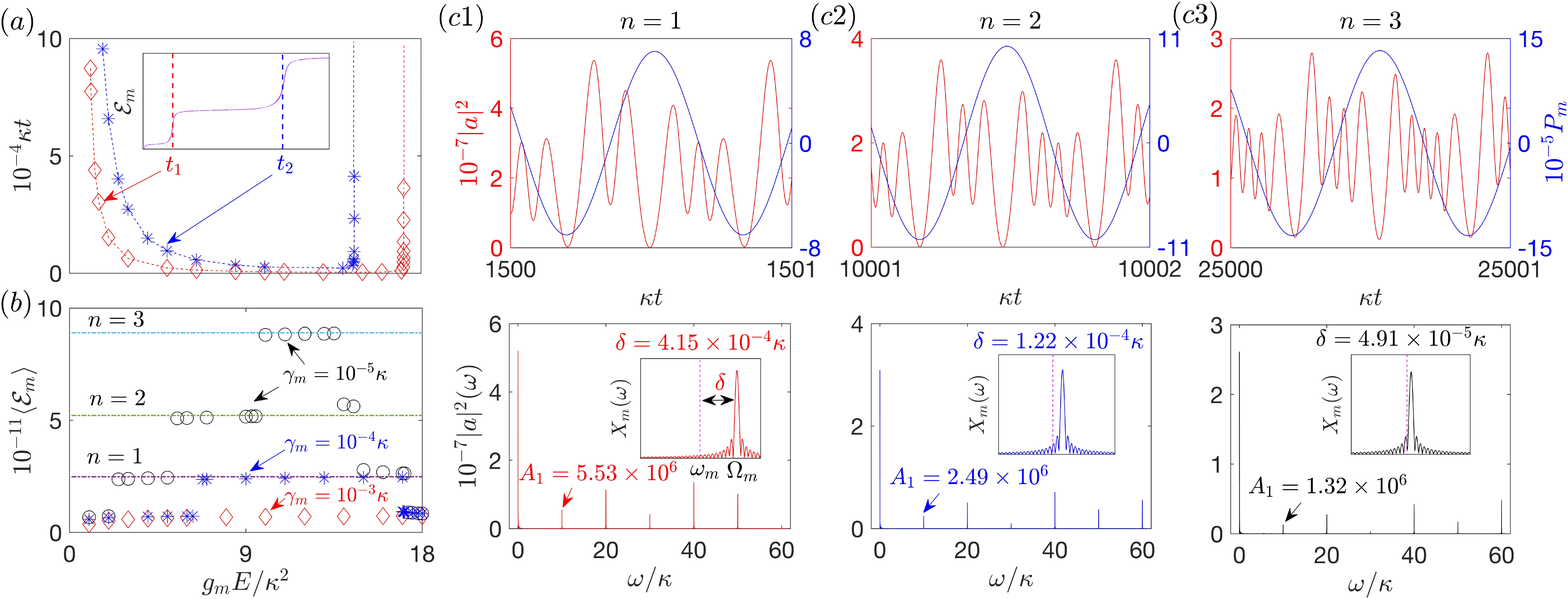}
	\caption{(a) The adjustable moment $t_1$, corresponding to the driving pulse's peak, for an ideal system ($\gamma_m=0$) to jump up to the step $n=1$, together with the adjustable $t_2$ to the next $n=2$, when this system of $g_m=10^{-4}\kappa$ and $\omega_m=10\kappa$ is driven at the RPB. (b) The distribution of stabilized average energy $\langle\mathcal{E}_m\rangle$ (over one oscillation period) after adding various mechanical damping rates $\gamma_m$. The indicated levels are the average positions of the energy steps in (a), and are extremely close to those defined in Fig. \ref{fig4}. (a) and (b) share the same horizontal axis. (c1)-(c3) The correlated field pattern and mechanical oscillation in both time domain and frequency domain, exemplified with the system operated at $g_mE/\kappa^2=10$ in (a) but with $\gamma_m=10^{-5}\kappa$. On each step from $n=1$ to $n=3$, the metastable or stable $|a(t)|^2$ reaches its bottom value twice whenever the mechanical resonator arrives at its top speed in each oscillation period, forming a composite pattern of $2+3$, $3+4$, or $4+5$ peaks.}
     \label{fig3}
\end{figure*}

An ideal situation is frictionless mechanical resonator ($\gamma_m=0$); the whole coupled system stabilizes only under the field damping at a rate $\kappa$. Above the Hopf bifurcation point ($E\approx 1200\kappa$) at the RPB, such a system in Fig. \ref{fig1} will enter oscillations. In most range of $E$ it is a normal stabilization (NS) showing a relatively short transient period, like the two processes in Fig. \ref{fig1}(b). However, after the amplitude $E$ is switched to an in-between value as in Fig. \ref{fig1}(c), the system will undergo an anomalous stabilization (AS): the mechanical energy $\mathcal{E}_m(t)=\frac{1}{2}X_m^2(t)+\frac{1}{2}P_m^2(t)$ ($P_m=\dot{X}_m/\omega_m$) repeatedly jumps to a higher step, together with a field pattern change, after each action of a sequence 
of field pulses. The energy step heights are determined by the system's fabrication (mostly by the parameters $g_m$ and $\omega_m$).

The scenarios in Fig. \ref{fig1}(b) are outside the two boundaries of the AS regime illustrated in Fig. \ref{fig2}(a), which serves as a partial phase diagram at RPB. Although the pump power range between 
the boundaries of the AS regime is considerable to a specific COMS, the corresponding pump frequency range is narrow, e.g., within a window of $0.36\kappa$ at $(\omega_m/\kappa, g_mE/\kappa^2)=(10,10)$; see Fig. \ref{fig2}(b). The mechanical energy steps and associated field patterns become deformed around the ratio $\omega_m/\kappa=1$ [the purple point in the lower left corner of Fig. \ref{fig2}(a)], and the phenomenon of AS gradually disappears with the further decreased $\omega_m/\kappa<1$. If the ratio is tuned through fiber-cavity coupling to $\omega_m/\kappa=1.1$ for the microresonator in Ref. \cite{omcomb}, this remnant phenomenon of AS is observable with a pump power of $4.2$ mW and within a frequency window of $2\pi \times 4.5$ MHz.

One can freely adjust the time-lags between the pulses that push the mechanical oscillation to higher amplitudes. In Fig. \ref{fig3}(a) we show how to change the length of a mechanical energy step by the parameter $g_mE/\kappa^2$. For a specific COMS with fixed $g_m$, the pulses' emergence will be tremendously delayed if the pump power is close to both boundaries of the AS regime. It is due to the critical slowing-down near the bifurcation points. A realistic system ($\gamma_m\neq 0$) does not change the jump moments to the higher steps but will lower the gap, 
\begin{eqnarray}
&&\Delta \mathcal{E}_m(n)= \int_{\delta t_n}\{g_m|a(t)|^2P_m(t)-\gamma_m P^2_m(t)\}dt,
\end{eqnarray}
between the $(n-1)$-th and $n$-th step by the second term of the above, with $\delta t_n$ being the $n$-th pulse's duration. For example, the mechanical damping up to $\gamma_m=2.73\times 10^{-5}\kappa$ lowers the first step ($n=1$) of the ideal system in Fig. \ref{fig3}(a) by about $1.2\%$. With a loss larger than this amount, the next pulse leading to $n=2$ cannot be formed anymore and the resonator will stay on the former forever. A more general result is in Fig. \ref{fig3}(b): any realistic system operated in the AS regime stabilizes near one of a series of fixed average energy $\langle\mathcal{E}_m\rangle$ according to its mechanical damping rate $\gamma_m$ and the optomechanical strength $g_mE/\kappa^2$, except for a sufficiently high $\gamma_m$ that only leads to the dynamical processes below the lower boundary of the AS regime. 

\begin{figure}[t]
	\centering\includegraphics[width=\linewidth]{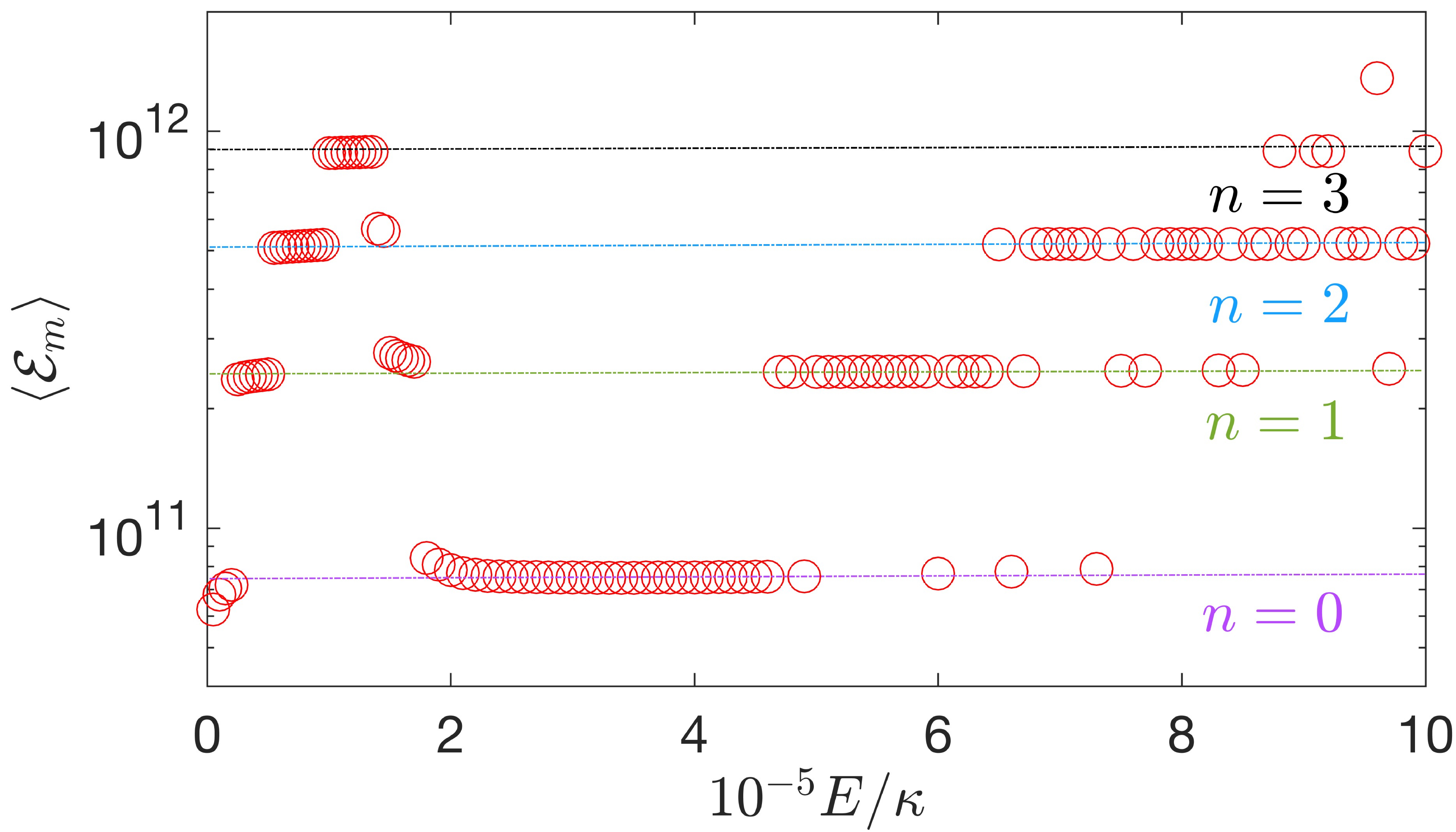}
	\caption{The stabilized average mechanical energy distribution for a COMS driven at the RPB and from the static initial condition. From the left (AS regime) to the right (NS regime), the system is locked to the orbits from $n=0$ to $n=3$. The stray point in the upper right corner is on $n=4$. Here, $g_m=10^{-4}\kappa$, $\omega_m=10\kappa$, and $\gamma_m=10^{-5}\kappa$.}  
	\label{fig4}
\end{figure}

Before finally stabilizing on the step $n=3$, a system with $\gamma_m=10^{-5}\kappa$ and operated at $g_mE=10\kappa^2$ temporarily stayed in two metastable states on $n=1$ and $n=2$ successively, having their corresponding cavity field patterns in Figs. \ref{fig3}(c1)-3(c3). Two factors explain why the system can be in these metastable and stable states during its time evolution. One is optical spring effect which shifts the actual frequency $\Omega_m$ of the mechanical oscillation
$X_m(t)=\sqrt{2\langle\mathcal{E}_m\rangle}\cos(\Omega_mt)+d$ ($d/\sqrt{2\langle\mathcal{E}_m\rangle}\ll 1$) 
by an amount $\delta=\Omega_m-\omega_m$. The other is the first sideband magnitude $A_1=2|\sum_{n=-\infty}^{+\infty}a^{\ast}_na_{n+1}|$ of the field intensity or intracavity photon number
\begin{eqnarray}
|a(t)|^2&=&\sum_{n=0}^{\infty}A_n\cos(n\Omega_m t+\phi_n),
\label{field}
\end{eqnarray}
where $a(t)=\sum_{n=-\infty}^{+\infty} a_n e^{i(\varphi(t)+n\Omega_mt)}$. Optical spring effect was regarded as a contribution to the effective spring constant $k_{eff}=m\omega_m^2-\frac{m\omega_m}{\hbar}V^{\prime\prime}(X_0)$ by the radiation potential at the equilibrium position $X_0$ of the mechanical resonator, and there is an analytical form of $\delta$ based on the approximated linear response of COMS \cite{rev2}. As a matter of fact, the mechanical resonator is generally under complicated radiation force during one oscillation period (see Fig. S-1 in Appendix II), significantly deviating from the linear approximation. Instead, the shift $\delta$ can be exactly read from the Fourier transform of a numerically simulated $X_m(t)$ as in Figs. \ref{fig3}(c1)-3(c3). The sideband magnitude $A_1$, which can be obtained from the Fourier transform of $|a(t)|^2$, determines the energy 
\begin{eqnarray}
	\langle\mathcal{E}_m\rangle&=&\frac{(g_m\omega_mA_1)^2}{2\delta^2(2\omega_m+\delta)^2+2\gamma_m^2(\omega_m+\delta)^2}
\label{energy}
\end{eqnarray}
from Eq. (2). Generally one has $\delta\neq 0$, so the corresponding mechanical amplitude can be permanently or temporarily stable even for an ideal system with $\gamma_m=0$.
From Fig. \ref{fig3}(c1) to 3(c3), the sideband magnitude $A_1$ dwindles after each jump of the mechanical amplitude. What compensates for the smaller driving force on a higher orbit is a further shrunk $\delta$, so that the energy according to Eq. (\ref{energy}) will be the one on the higher orbit. The jointly reduced $\delta$ and $A_1$ in Fig. \ref{fig3}(c2) give $\langle\mathcal{E}_m\rangle=5.21\times 10^{11}$, well close to $5.19\times 10^{11}$ on the level $n=2$ in Fig. \ref{fig4}. In Fig. 3(c3) the system has reached a completely stable optomechanical oscillation after the mechanical frequency shift lowers to $\delta\sim \gamma_m$. More examples are two different $E=2.5\times 10^4\kappa$ and $E=1.7\times 10^5\kappa$, under both of which the COMS in Fig. \ref{fig4} finally stabilizes on $n=1$. Their corresponding $(\delta, A_1)=(2.48\times 10^{-5}\kappa, 3.373\times 10^5)$ and $(3.45\times 10^{-3}\kappa, 4.996\times 10^7)$ lead to the $\langle\mathcal{E}_m\rangle$ adequately near $2.476\times 10^{11}$ on the level $n=1$. Through an automatic adjustment of $\delta$ with $A_1$, a system operating in the AS regime is always locked close to one of the reference energy levels like in Fig. \ref{fig4}. 

The cavity field corresponding to the $n$-th locked orbit has a special pattern of $(n+1)$ plus $(n+2)$ peaks during each mechanical oscillation period; see Figs. \ref{fig3}(c1)-3(c3). In contrast, the patterns in some NS regimes are non-compound, not splitting into two distinct parts in each oscillation period, like the stabilized one in the upper panel of Fig. \ref{fig1}(b). A field oscillation or its spectrum can be used to identify the associated mechanical orbit.

After the pump near the RPB becomes more powerful, the system will cross the upper boundary of the AS regime and be locked again to the more regularly distributed orbits (the levels $n=1,2,3$ in Fig. \ref{fig4} are according to the stabilized $\langle\mathcal{E}_m\rangle$ in this regime). Given the experimental setup in Ref. \cite{omcomb}, the upper boundary at $\omega_m/\kappa=10$ can be reached with a pump power around $55.5$ mW. Now the system is in NS but the possible field patterns still assume the composite ones as those in Figs. \ref{fig3}(c1)-3(c3). The lower panel of Fig. \ref{fig1}(b) displays one more composite pattern of $1+2$ peaks, corresponding to the lowest orbit $n=0$. Those locked oscillations in the right part of Fig. \ref{fig4} are also due to the correlated $A_1$ and $\delta$ in Eq. (\ref{energy}). There exist random transitions between them under a slight change of the larger $E$ in this regime, and they are similar to those caused by a two-tone drive with its tones differed by the mechanical frequency $\omega_m$ \cite{two-tone1, two-tone2} (in that two-tone scenario a locked orbit similar to $n=0$ is realized within the range of $E$ for the AS regime).   

One relevant issue is about the multiple dynamical attractors of a system driven at the fixed $\Delta$ and $E$. The mechanical amplitudes at the attractors were determined by the balanced average power flows into and out of a mechanical resonator \cite{multi1, multi2, multi4}, and such discrete mechanical amplitudes thus change with pump detuning and power continuously. We approach the multistability by adding initial mechanical momentum to the dynamical processes described by Eqs. (1) and (2), so that the system can directly reach different basins of attraction. Apart from below the lower boundary of AS regime, where the attractors were observed by the signature of non-compound field patterns \cite{multi5}, we find that the attractors beside a RPB are totally locked to the fixed orbits like those in Fig. \ref{fig4} (see Appendix III). Given enough pump power, locked orbits can exist in the vicinity of $\Delta=n\omega_m$ ($n$ are integers). They exclude optomechanical chaos \cite{chaos1,chaos2,chaos3,chaos4,chaos4,chaos-r} to somewhere beyond the range of Fig. \ref{fig4}. Under lower drive powers, chaos only emerges at the detuning points without locked orbits, and 
one route towards this type of chaos is exemplified in Appendix IV. 

Although locked orbits can be encountered elsewhere over the detuning $\Delta$ of a pump laser, the phenomenon of AS is unique, only in a neighborhood of RPB. It is an effective channel to add up mechanical energy with lower pump powers. Not only the oscillations as phonon lasers \cite{pt1,pt4,pt5} can be excited, coupling one COMS to other cavities will also preserve the phenomenon of AS, as long as the coupling intensities are below certain limit. It is possible for such coupled systems operating in the AS regime to outperform the $\mathcal{PT}$-symmetric phonon lasers \cite{pt2,pt3,pt6} aided with optical gain.

Based on the full nonlinear dynamics well depicted by Eqs. (1) and (2), we have demonstrated that the dynamical behaviors of a COMS under single-tone drives are much richer than what were thought in the past, particularly near its RPB where only a Hopf bifurcation was known. In fact, responding to a continuously enhanced pump laser driving at the RPB, the system will first enter one more regime of AS and then another regime of NS. Like a celestial body deformation in tidal locking, a metastable or stable mechanical oscillation in these regimes is deformed by proper frequency shift, to correlate with the corresponding field sidebands, so that it can be locked to one of the reference energy levels as in Fig. \ref{fig4}, rather than its monotonous amplitude increase with the pump power higher than the Hopf threshold. The current understandings of optomechanical dynamics, especially the existence of AS processes, will find applications in the development of the relevant setups.

This research was supported by ANID Fondecyt Regular de Chile (1221250) and Natural Science Foundation of China (Grant Nos. 12374348, 12293054, 12341403).

\renewcommand{\theequation}{S-\arabic{equation}}
\setcounter{equation}{0}  

\renewcommand{\thefigure}{S-\arabic{figure}}
\setcounter{figure}{0}  

\renewcommand{\thetable}{S-\arabic{table}}
\setcounter{table}{0}  

\section*{Appendix I. System Dynamics}

We first provide an explanation of the notations used in the numerical simulations.
A cavity optomechanical system (COMS) \cite{rev1,rev2} is modeled by two coupled oscillators; one is for the mechanical mode with the intrinsic frequency $\omega_m$ and the other represents the cavity field of the original frequency $\omega_c(0)=cN\pi/L$ with $N$ being an integer and $L$ the cavity size (this argument with the example of F-P cavity applies to all setups with the similar topology). We define the dimensionless displacement $X_m$ and the dimensionless $P_m$ of the mechanical oscillator through the relations to the real diaplacement and momentum: $x_m=\sqrt{\frac{\hbar}{m\omega_m}}X_m$ and $p_m=\sqrt{m\hbar\omega_m}P_m$, so the mechanical energy $(1/2)m\omega_m^2x_m^2+(1/2)p_m^2/m$ is reformulated as $\hbar\omega_m\times (1/2)(X_m^2+P_m^2)=\hbar\omega_m\mathcal{E}_m$, where the dimensionless energy $\mathcal{E}_m$ is equivalent to a phonon number. Under a radiation pressure the resonant cavity frequency becomes
\begin{eqnarray}
	\omega_c(x_m)&=&\frac{cN\pi}{L+x_m}=\omega_c(0)+\frac{d \omega_c}{d x_m}(0)~x_m+\cdots\nonumber\\
	&=&\omega_c(0)-g_m X_m+\cdots
	\label{1}
\end{eqnarray}
after the mechanical displacement $x_m$, where $g_m=(x_{ZPF}/L)\omega_c(0)$ and $x_{ZPF}=\sqrt{\frac{\hbar}{m\omega_m}}$. Given $x_{ZPF}\sim 0.1$ fm and $L\sim 10$  $\mu$m of an experimental setup \cite{omcomb}, for example, the mechanical amplitude even when $\mathcal{E}_m=10^{12}$ is within $150$ pm (less than three Bohr radius), validating the above first-order expansion. The field energy is $\hbar\omega_c(x_m)|a|^2$, where $|a|^2$ is the intracavity photon number. The cavity has an intrinsic loss rate $\kappa_i$, while the mechanical oscillator damps at the rate $\gamma_m$.

\begin{figure*}[t]
	\centering\includegraphics[width=\linewidth]{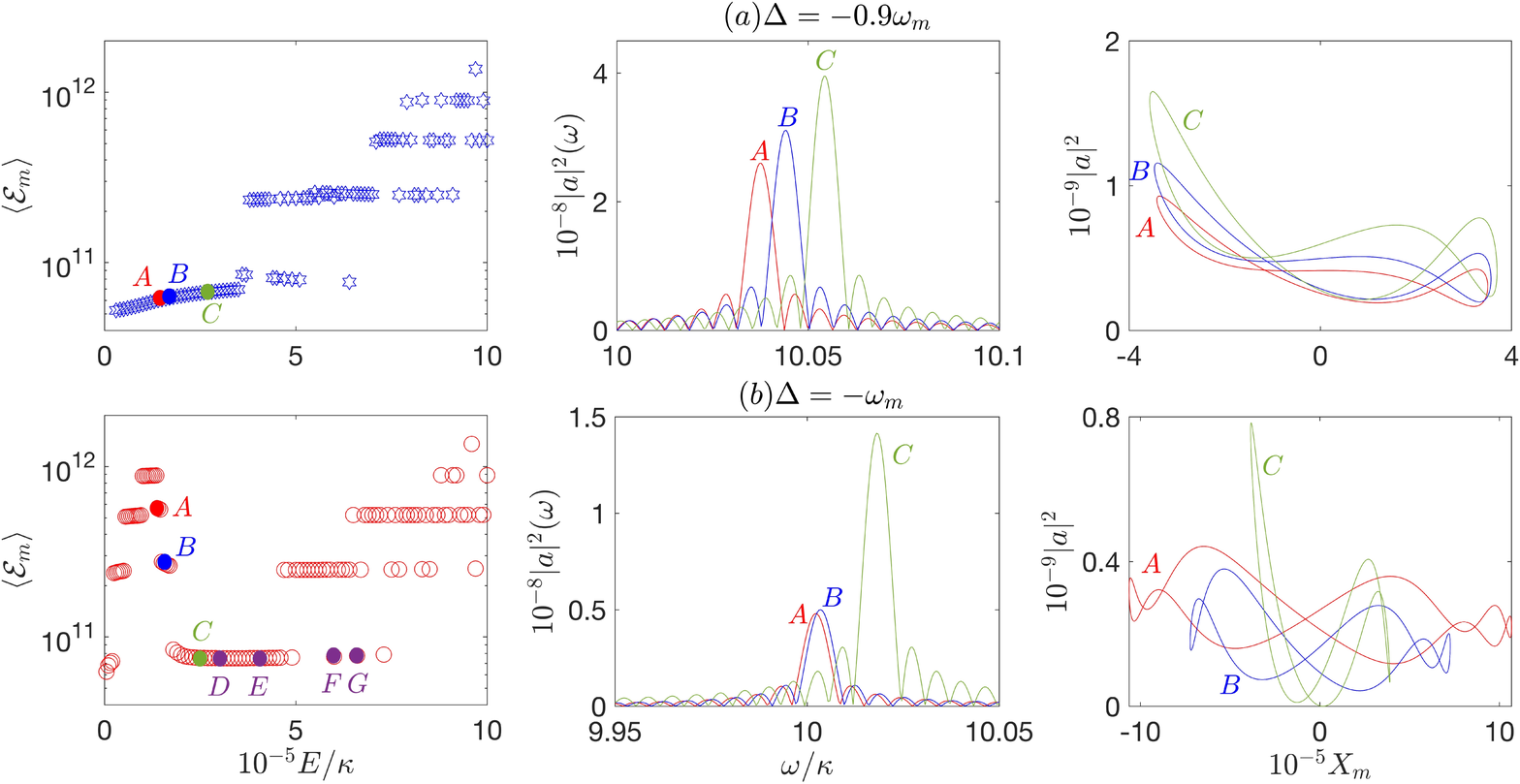}
	\caption{Comparisons of the average mechanical energy in the final stability, the shift of the first 
		sideband (with its magnitude $A_1$) of the field intensity $|a(t)|^2$ at the indicated points, as well as the corresponding field intensity (proportional to the optomechanical force) over one mechanical oscillation period. (a) The results when the system is operated out of the AS regime. The locked mechanical orbits still exist if the drive power becomes sufficiently high, and the middle and right show the properties at the indicated points in the left panel. (b) The corresponding behaviors in the AS regime. Here the system parameteres are the same as those in Fig. 4 of the main text.}
	\label{figs1}
\end{figure*}

\begin{figure}[t]
	\centering\includegraphics[width=\linewidth]{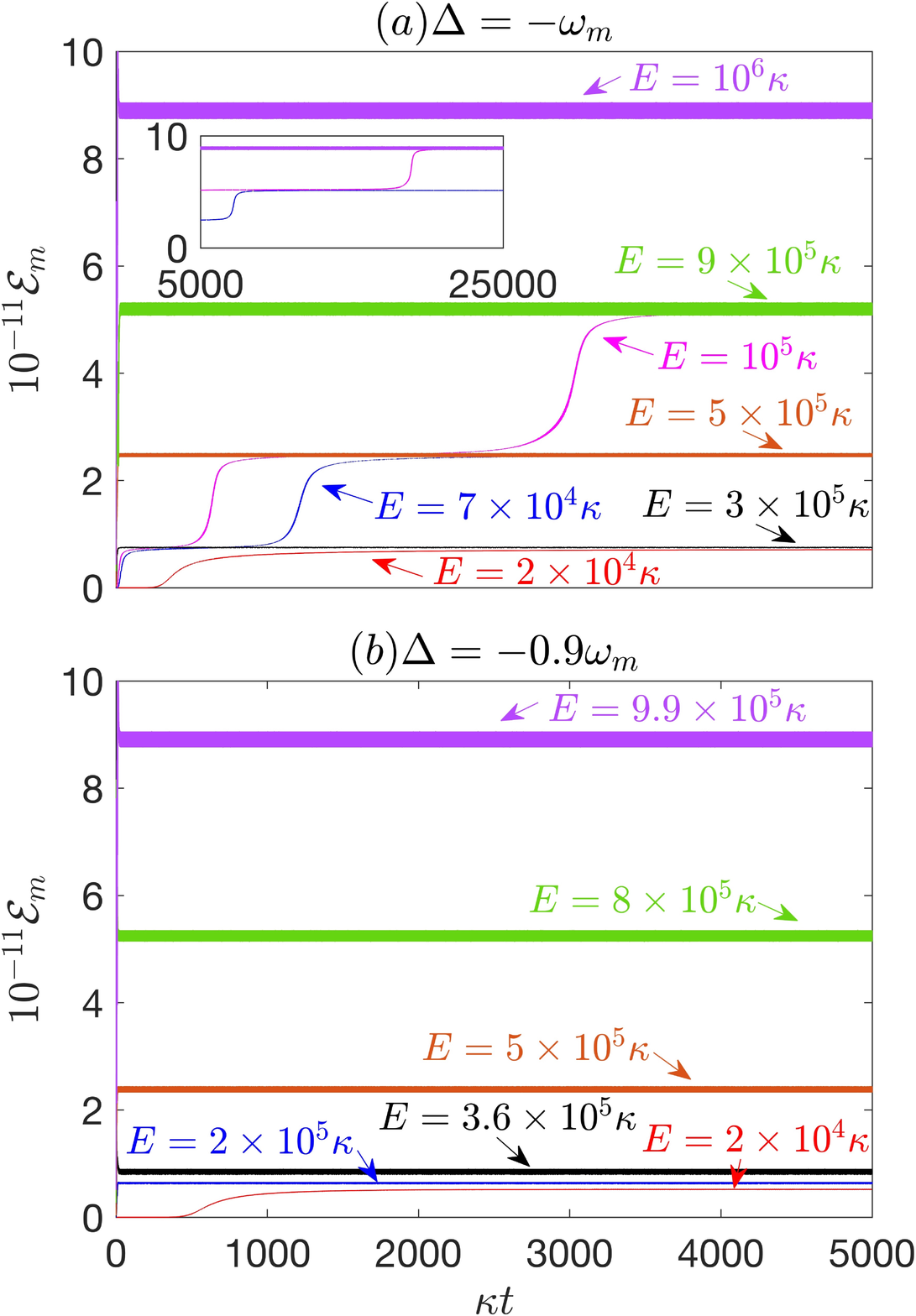}
	\caption{Sample evolution processes corresponding to what are described in Fig. \ref{figs1}. (a) In the AS regime, all evolution courses (except for the lowest one) meet with one another in the fixed vertical positions during different periods of time. The two drive amplitudes $E=7\times 10^4\kappa$ and $E=10^5\kappa$ give rise to the AS phenomenon with the corresponding mechanical energy jumping to a higher step from time to time. (b) Out of the AS regime, the courses due to the lower $E$ distribute continuously, but the four courses due to $E=3.6\times 10^5\kappa$, $E=5\times 10^5\kappa$, $E=8\times 10^5\kappa$ and $E=9.9\times 10^5\kappa$ are still locked to the orbits $n=0$, $1$, $2$, and $3$, respectively. }  
	\label{figs2}
\end{figure}

\begin{figure}[t]
	\centering\includegraphics[width=\linewidth]{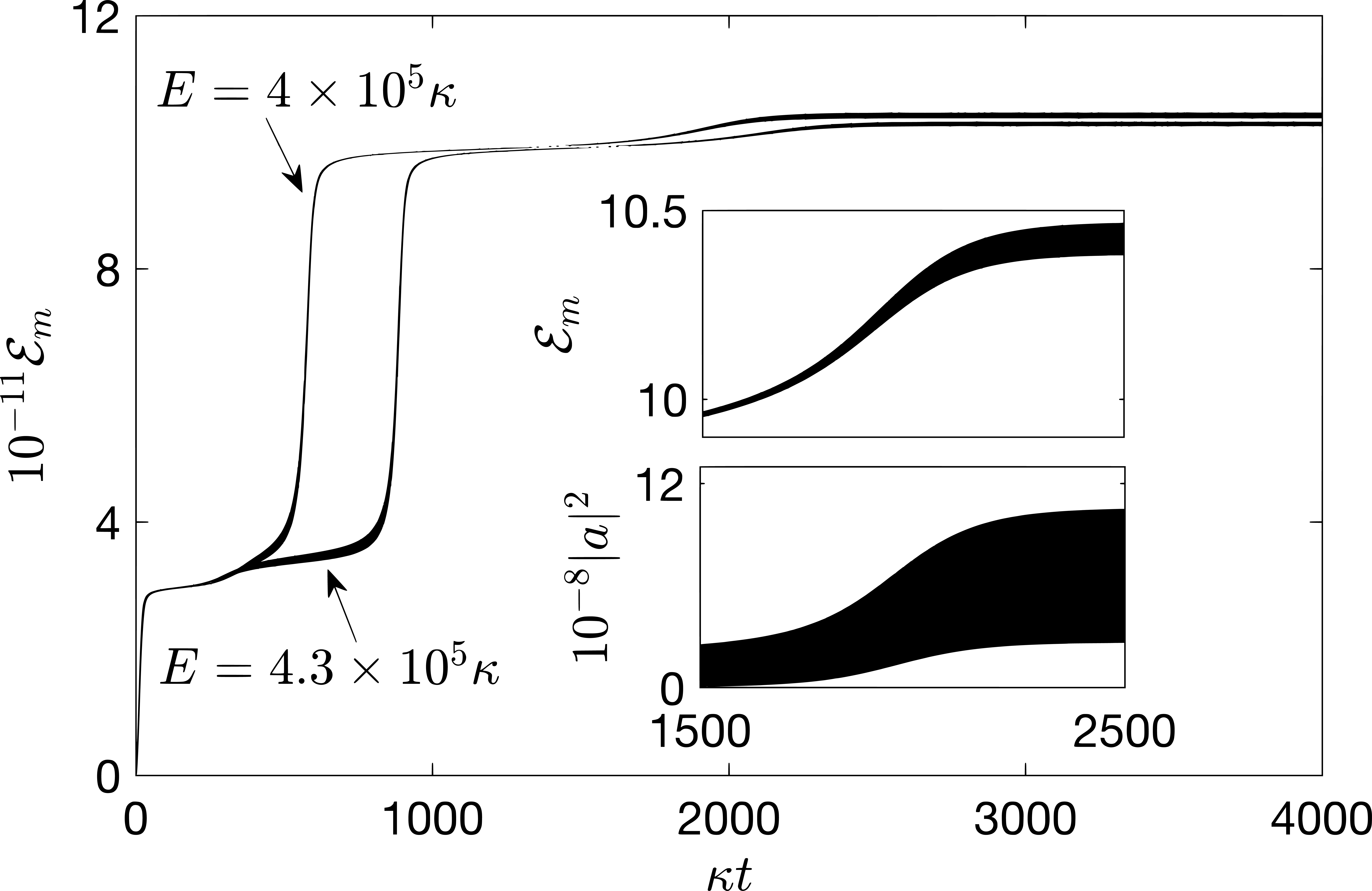}
	\caption{One example of a two-mode squeezing effect in the AS regime. The insets show the simultaneous increase of the mechanical energy and the cavity field intensity, due to the squeezing effect at the drive amplitude $E=4\times 10^5\kappa$. Here the intrinsic mechanical frequency is increased to $\omega_m=20\kappa$ as compared to the previous examples. It takes place near the upper boundary of the AS regime shown in Fig. 2(a) of the main text. The position of the standard orbit of locked is marked with $n=1$. }  
	\label{figs3}
\end{figure}

Once a pump laser with the frequency $\omega_l$ and power $P$ drives this coupled system, the dynamical process after considering the first order correction to the cavity field energy is described by the following equations:
\begin{eqnarray}
	&&\dot{a}=-\kappa a-i\omega_c a+ig_mX_ma+Ee^{-i\omega_l t},\nonumber\\
	&&\dot{X}_m=\omega_mP_m,\nonumber\\
	&&\dot{P}_m=-\gamma_mP_m-\omega_mX_m+g_m|a|^2,
	\label{2}
\end{eqnarray}
where the drive amplitude $E=\sqrt{2\kappa_eP/(\hbar \omega_l)}$ is related to the pump-cavity coupling 
rate $\kappa_e$ so that $\kappa=\kappa_e+\kappa_i$. Here we neglect the associated noise terms because they do not affect the concerned processes, though they can be numerically modeled by stochastic functions \cite{os8}. By a global rotation $a\rightarrow ae^{-i\omega_c t}$ ($\omega_c(0)\equiv \omega_c$), the cavity field part of the above equations will be transformed to a more compact form
\begin{eqnarray}
	&&\dot{a}=-\kappa a+ig_mX_ma+Ee^{i\Delta t},
	\label{3}
\end{eqnarray}
where the detuning $\Delta=\omega_c-\omega_l$ is defined as in a review article \cite{rev1} (among some other literature, including another review article \cite{rev2}, the detuning differs by an opposite sign). This is Eq. (1) in the main text, and the mechanical part of Eq. (\ref{2}) can be combined to a second order differential equation as Eq. (2) in the main text. By means of another transform $a\rightarrow ae^{-i\omega_l t}$, the equation of the field part takes the form 
\begin{eqnarray}
	&&\dot{a}=-\kappa a-i\Delta a+ig_mX_ma+E
	\label{4}
\end{eqnarray}
with a time-independent drive. The calculated intracavity photon number $|a(t)|^2$ is independent of the used reference frame due to the above-mentioned field phase transformations. In the numerical calculations we use the dimensionless system parameters scaled by $\kappa$ so that the system evolves with the dimensionless time $\kappa t$, and it is the reason for not adopting another form of the damping rates as $\kappa/2$ and $\gamma_m/2$ in the dynamical equations, which is seen in some other literature. 

The Hopf bifurcation between static equilibrium and optomechanical oscillation (self-induced or backaction-induced oscillation named in Ref. \cite{rev2}) is usually from the view of the mechanical part. 
A direct numerical simulation of the mechanical motion above a Hopf bifurcation threshold clarifies that the actual frequency $\Omega_m$ of the stabilized oscillation, $X_m(t)=A\cos(\Omega_m t)+d$ (the oscillation phase is chosen to be zero with a reference time), is shifted from the intrinsic one $\omega_m$. This shift due to optical spring effect can be accurately read from the Fourier transform of $X_m(t)$ or the distance between the sidebands of the field intensity $|a(t)|^2$, where $a(t)=e^{i\varphi(t)}\sum_{n=-\infty}^{\infty}a_n e^{in\Omega_mt}$. 
If the system is operated in certain regimes determined by the parameters $\Delta$ and $E$, the realized field sidebands $a_n$ ($n\in Z$) and mechanical motion characterized by $\Omega_m$ and $A$ will become highly correlated: the first sideband of the driving force for the mechanical part in Eq. (\ref{2}), $g_mA_1\cos(\Omega_mt+\phi_1)$ where 
$A_1 =2|\sum_{n=-\infty}^{+\infty}a^{\ast}_na_{n+1}|$, which is the closest to the resonance of the mechanical oscillation, always drives the mechanical amplitude,
\begin{eqnarray}
	A=\frac{g_m\omega_mA_1}{\sqrt{(\omega_m^2-\Omega_m^2)^2+\gamma_m^2\Omega_m^2}},
	\label{6}
\end{eqnarray}
to near a series of fixed values. By Eq. (\ref{6}), the realized 
sideband magnitude $A_1$ and the correspondingly shifted mechanical oscillation frequency $\Omega_m$ should automatically adjust with each other, if the different drive amplitudes $E$ (at the same $\Delta$) can realize almost the same mechanical amplitude $A$ when they are applied to the same system. Through this type of correlation, a system under the same $E$ can also reach different metastable states in an anomalous stabilization (AS) process.

\begin{figure*}[t]
	\centering\includegraphics[width=\linewidth]{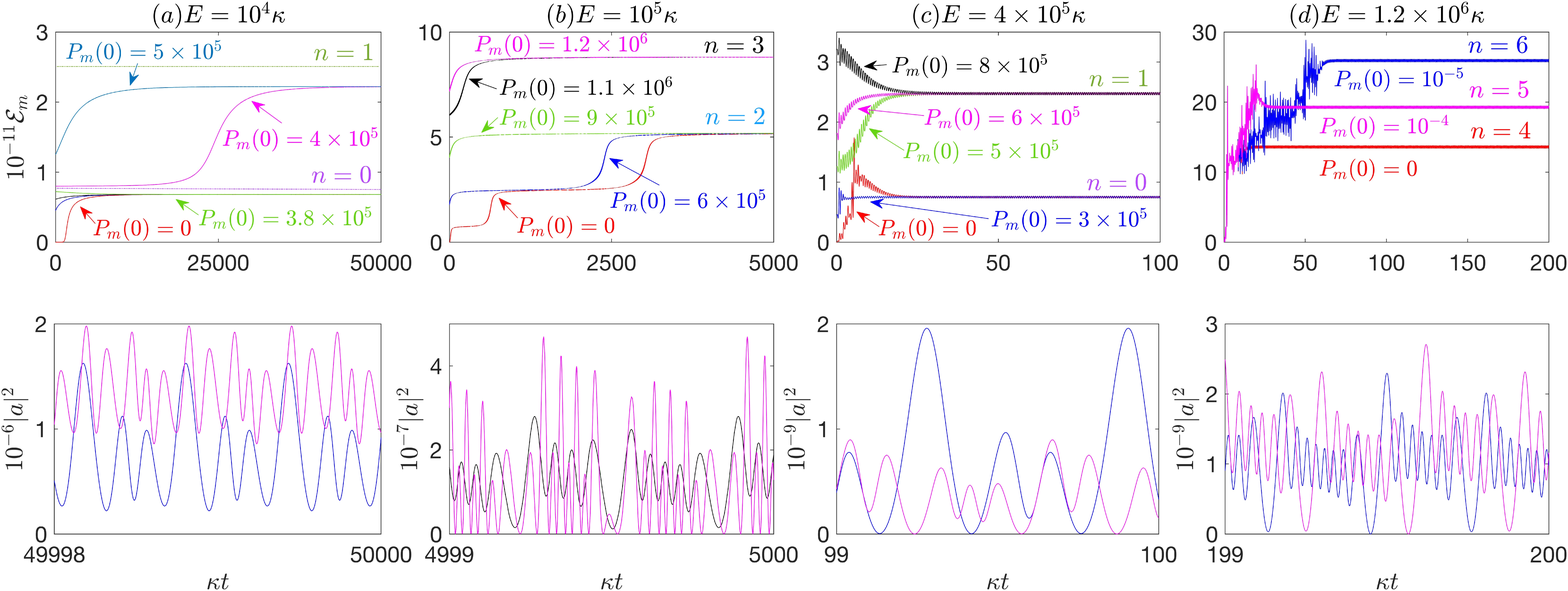}
	\caption{Influence of the initial momentum on the dynamical evolution of a system with its parameters as those in Fig. \ref{figs1}. From (a) to (d), the system driven at $\Delta=-\omega_m$ goes through four different regimes: unlocked orbits, locked orbits in the AS regime, locked orbits to lower energy levels, and locked orbits to higher energy levels. In the regimes of locked orbits, the evolved orbit can be identified by a composite field pattern. With a magnified view, the non-compound patterns in the lower panel of (a) will be seen more clearly to reach their bottom values only once in each mechanical oscillation period. The lower panel of (d) displays the composite patterns of $6+7$ peaks and $7+8$ peaks, which are characteristic of the $5$-th and $6$-th locked orbit. }
	\label{figs4}
\end{figure*}

\section*{Appendix II. More Details of Field-Oscillator Correlation}

The collaboration between the field sidebands and the mechanical frequency shift can be better seen from comparing the associated dynamical behaviors within and without the AS regime discussed in the main text. If the drive laser frequency is tuned to the point $\Delta=-0.9\omega_m$ as in Fig. \ref{figs1}(a), which is out of the AS regime, there is a continuously increasing period of the stabilized average mechanical energy $\langle\mathcal{E}_m\rangle$ before the pump power $P=(1/2)\hbar\omega_l E^2/\kappa_e$ reaches a regime of locked orbits. Taking three sample points A, B and C in this range, one finds that their corresponding mechanical frequency shifts $\delta=\Omega_m-\omega_m$ are considerable, as demonstrated by the shifts of the first field sideband from the position $\omega=\omega_m=10\kappa$ in Fig. \ref{figs1}(a). On the other hand, if the system is driven at the point $\Delta=-\omega_m$ within the regime of AS, the magnitudes of the first field intensity sideband at the corresponding operation points become much less as shown in the middle panel of Fig. \ref{figs1}(b), but the stabilized mechanical energy $\langle\mathcal{E}_m\rangle$ at the corresponding points A and B are much higher. The reason is that the mechanical frequency shifts at these points become much smaller, so the first sideband of the cavity field force, with a weaker magnitude proportional to $A_1$, goes much closer to the mechanical resonance and can therefore realize a much larger mechanical oscillation amplitude. We also take four more sample points D, E, F, and G in the regime of locked mechanical oscillations under still higher drive powers, to calculate their mechanical amplitudes with the first sideband amplitude $A_1$ and mechanical frequency shift $\delta$. These results are in good agreement with the mechanical amplitudes directly read from the simulated $X_m(t)$; compare the calculated $A_C$ and the directly read $A_R$ in Table S-1. 

\begin{table}[h!]
	\caption{\label{tab:example} The drive amplitudes on the first row are those from point D to point G in Fig. \ref{figs1}(b).} 
	\begin{ruledtabular} \begin{tabular}{lllll} 
			E & $3\times 10^5\kappa$ &$4\times 10^5\kappa$ & $6\times 10^5\kappa$ & $6.6\times 10^5\kappa$\\
			$A_0$	 &$7.42\times 10^8$ &$1.30\times 10^9$ &$2.97\times 10^9$ &$3.64\times 10^9$ \\ 
			$A_1$	 &$2.02\times 10^8$ &$3.70\times 10^8$ &$9.41\times 10^8$ & $1.198\times 10^9$\\
			$\delta$	 & $0.026\kappa$ & $0.048\kappa$ & $0.1206\kappa$ & $0.1524\kappa$\\
			$A_C$ 	 & $3.88\times 10^5$& $3.86\times 10^5$ & $4.01\times 10^5$ &$4.02\times 10^5$\\
			$A_R$ 	 &$3.91\times 10^5$ & $3.90\times 10^5$ &$3.89\times 10^5$ & $3.90\times 10^5$\\
		\end{tabular} 
	\end{ruledtabular} 
\end{table}

Another direct way to see the effect of such field-oscillator correlation is through the time evolution processes corresponding to those in Fig. \ref{figs1}. For example, the time evolution of mechanical energy under the drive $E=10^5\kappa$ in Fig. \ref{figs2}(a) goes through a number of metastable states until it finally stays on the orbit $n=3$. During each stage of time evolution, the sideband amplitude $A_1$ and the frequency shift $\delta$ always collaborate to give the proper mechanical amplitude $A$ on the present orbit. The cavity field and mechanical oscillation are explicitly related only through Eq. (\ref{6}), but a miraculous fact is that the evolution processes under all drive amplitudes $E$ (except for the much lower ones) must be on a series of fixed mechanical orbits, no matter whether the mechanical oscillator stays on them permanently or temporarily. All evolution courses due to the different drive amplitudes $E$ (except for the lowest one) in Fig. \ref{figs2}(a) meet one another during different periods of time, but they only metastably or stably locate on the displayed positions of $\mathcal{E}_m$ without settling down in the middle between any two of the displayed mechanical energy values. In Fig. \ref{figs1}(a) and Fig. \ref{figs2}(b), however, the stabilized orbits distribute continuously in much wider range of the drive amplitude $E$, with the discrete ones in the AS regime being replaced by them.

One side issue is about the exact form of the mechanical frequency shift $\delta=\Omega_m-\omega_m$. According to a linearized dynamics based on the static steady states of COMS, the shift can be derived to the analytical form (the notations are translated into those used in the current work) \cite{rev2}:
\begin{eqnarray}
	\delta(\Delta=-\omega_m)=g_m^2A_0\frac{2\omega_m}{\kappa^2+4\omega_m^2},
	\label{7}
\end{eqnarray}
where $A_0=\sum_{k=-\infty}^{\infty}|a_k|^2$ is the zeroth sideband as the average photon number in cavity. With the $A_0$ obtained from the Fourier transform of the field intensity $|a(t)|^2$ (see Table S-1), the shift at point D in Fig. \ref{figs1}(b) is calculated according to the above formula as $0.37\kappa$, and another shift calculated with Eq. (\ref{7}) for point G is $1.815\kappa$, much larger than the real ones read from numerical calculations. The increased sideband $A_0$ by a higher pump power obviously increases the static displacement $d$ of the mechanical oscillator, so the oscillation amplitude $Ad$ of the energy $\mathcal{E}_m(t)$ become obvious to the evolution courses due to the higher $E$, those thick trajectories in Figs. \ref{figs2}(a) and \ref{figs2}(b). However, the influence of the sidebands on mechanical oscillation frequency is much more complex. As seen from the right column in Fig. \ref{figs1}, the mechanical oscillator experiences very complicated optomechanical force proportional to $|a(t)|^2$ during each mechanical oscillation period, so the system will definitely go beyond the linear response used in Ref. \cite{rev2}. Therefore, it is much more accurate to find the mechanical frequency shift by a numerical simulation of $X_m(t)$, which is dominated by only one frequency component except in the regimes near chaos. 

Another issue is where to see the two-mode squeezing effect highlighted by the linearized dynamics of COMS. The squeezing effect leads to a simultaneous growth of the field intensity and mechanical amplitude. One of its regimes is in where the continuously distributed orbits locate in Fig. \ref{figs1}(a); in Fig. \ref{figs1}(b) with the existence of AS, the range of this regime is rather short at the low drive powers. In the regimes of locked orbits, however, the squeezing effect manifests in very limited locations. One example is in Fig. \ref{figs3}, where the drive amplitude $E$ is close to the upper boundary of the AS regime, which is bent upwards due to a larger $\omega_m$ (see Fig. 2(a) in the main text). The correction from this squeezing effect only slightly modifies a locked orbit. It indicates that the locking mechanism dominates over the squeezing effect in this regime, and the dynamics of COMS cannot be linearized there.

\section*{Appendix III. Multistability in Different Regimes}

The discussions in elsewhere assume that the system evolves from the initial condition, $(X_m,P_m)=(0,0)$ and $a=0$. Other initial conditions can surely affect the evolution process according to Eq. (\ref{2}). Previously, this problem was approached by a reduced dynamical picture of reaching the possible equilibria of the averaged energy flows into and out of the mechanical oscillator, to solve the equation $G(A, E, \Delta)\equiv\gamma_m+\Gamma_{opt}(A)=0$ \cite{rev2}, where $\Gamma_{opt}(A)=-\langle F\dot{x}_m\rangle/(m\langle\dot{x}_m^2\rangle)$, with $F(t)=g_m|a(t)|^2$, is an effective mechanical damping rate. This equation determines the mechanical oscillation amplitudes $A$ at the dynamical attractors in the space of pump detuning \cite{rev2,multi1}. Its solution, as a set of discrete mechanical amplitudes $A$, changes with pump detuning and pump power in a continuous way. It can be seen from the solution, $a(t)=e^{i\phi(t)}\sum_{n=-\infty}^{\infty}a_n e^{in\Omega_mt}$, to Eq. (\ref{4}) 
that the sideband amplitudes, 
\begin{eqnarray}
	a_n=E\frac{J_n(-g_mA/\Omega_m)}{in\Omega_m+\kappa+i(\Delta-g_md)},
	\label{c}
\end{eqnarray}
in terms of the Bessel functions $J_n(x)$ (after the notations are translated into those in the current work) carry the variables $\Delta$ and $E=\sqrt{2\kappa_eP/(\hbar \omega_l)}$. Generally, one has 
\begin{eqnarray}
	\frac{dA}{dE} &=-\frac{\partial G/\partial E}{\partial G/\partial A}\neq 0,
\end{eqnarray}
given a fixed detuning $\Delta$, in contrast to the almost fixed $A$ on a locked orbit with $\langle\mathcal{E}_m\rangle\approx (1/2)A^2$.

\begin{figure}[b]
	\vspace{0.3cm}
	\centering\includegraphics[width=\linewidth]{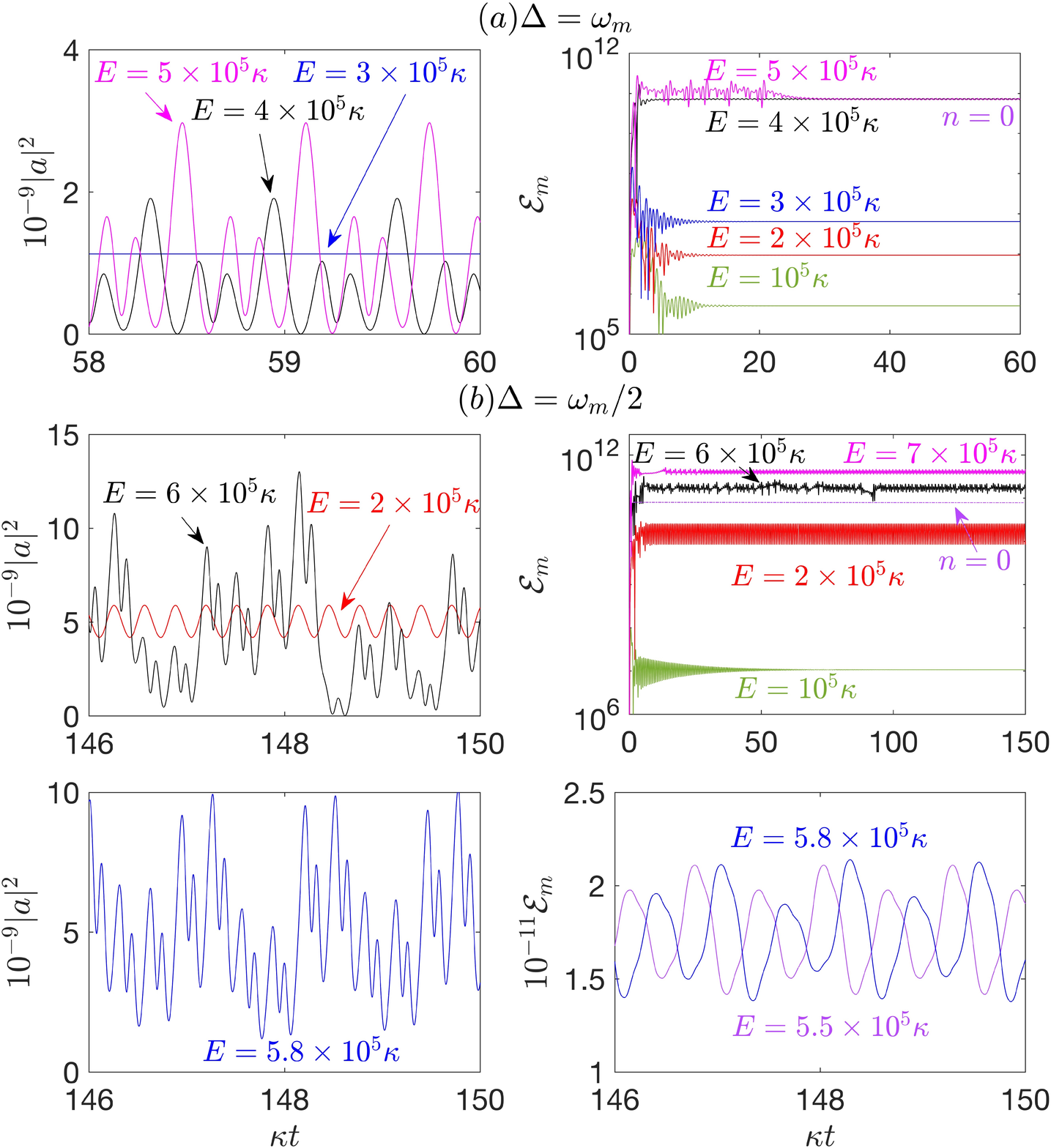}
	\caption{Dynamical evolution processes in the regimes of red detuning. (a) The system is driven at the red-detuned point $\Delta=\omega_m$. In this scenario the system can stabilize in static steady states (with $E\leq 3\times 10^5\kappa$ here), which are applied for optomechanical cooling. After the pump power is enhanced further, the system will stabilize to the locked orbits. (b) The detuning point is changed to $\Delta=0.5\omega_m$. Chaos and a transition to chaos emerge, by the drive powers that realize the locked orbits at $\Delta=\pm\omega_m$. The black curves at $E=6\times 10^5\kappa$ show the aperiodic field intensity and mechanical energy exactly in chaos. In the lower panel, the existence of fractional sidebands makes more complicated quasi-periodic oscillation patterns, as a signature of the transition towards a chaos. The system parameters are the same as those in Fig. 4 of the main text.}  
	\label{figs5}
\end{figure}

Here we study the dynamical evolution processes by directly adding initial momentum to mechanical oscillator, which is a possible way to perturb the system. We focus on the attractors at the resonance point of blue detuning ($\Delta=-\omega_m$) and find that they vary in four different regimes determined by the drive power. Below the lower boundary in the phase diagram, Fig. 2(a) in the main text, the attractors are reached as in Fig. \ref{figs4}(a). In this regime the attractors distribute discretely; with a considerable initial energy, the system will still evolve to the same orbit as with the initial condition $(X_m,P_m)=(0,0)$ and $a=0$. These discrete orbits should not be mixed up with the locked ones at the fixed positions (compare them with the indicated positions of the locked orbits). If the drive amplitude $E$ is changed to another one, their positions will be changed too, but the positions of those locked orbits are independent of the drive power. More examples in this regime showing the non-compound intracavity field patterns have been studied with a system of $\omega_m=33.2\kappa$ and $\gamma_m=1.99\times 10^{-4}\kappa$ (after the translation into our notations) \cite{multi4}, for which the lower boundary of the AS regime is much higher than the one in Fig. 2(a) of the main text. After the pump power is increased to above the lower boundary in Fig. 2(a) of the main text, the possibly reachable attractors will be only those locked orbits, as seen from Fig. \ref{figs4}(b), but the stabilized field pattern can be modified by initial condition [compare the two curves in the lower panel of Fig. \ref{figs4}(b)]. The multistability above the upper boundary in Fig. 2(a) of the main text becomes more varied. In somewhere the system can only evolve to the low orbits as in Fig. \ref{figs4}(c), a different attractor among the locked ones should be reached only after changing a considerable amount of the initial energy. The stabilized field patterns are the composite ones that are characteristic of the locked orbits. However, under still stronger drives that bring the system to high orbits, a tiny change of the initial condition will lead to another orbit; see Fig. \ref{figs4}(d). This feature that is similar to one characteristic of chaos exists to the high orbits as limit cycles. 

In the three regimes of locked orbits, which must be found by the coupled equations, Eq. (\ref{2}), the system can go from one fixed attractor to another if the drive power is changed and/or another initial condition is adopted, but the corresponding mechanical oscillation amplitudes of the attractors never have a considerable variation. In contrast, given the fixed $\gamma_m$ of a system, the attractors' amplitudes $A$ determined by the equation $\gamma_m+\Gamma_{opt}(A)=0$ \cite{rev2,multi1,multi4} will change with the drive amplitude $E$. Therefore, the two predictions for the dynamical attractors can get close only in the regime of continuously distributed stable states, below the lower boundary of the AS regime. 

\section*{Appendix IV. Connection with Chaos}
Chaos emerges at many locations in the parameter space of COMS. In Ref. \cite{os9} a rough distribution of optomechanical chaos is shown above the pump powers realizing the regimes of limit cycles, where actually exist even richer phenomena such as the locked mechanical orbits. We here show that optomechanical chaos actually hides between stabilized oscillations at some detuning points $\Delta$ for the pump to drive a COMS. 

At the special detuning points $\Delta=n\omega_m$ ($n\in Z$), a COMS can be locked to the fixed orbits if the driving laser power becomes sufficiently high. One more example is at the point of red detuning $\Delta=\omega_m$; see Fig. \ref{figs5}(a). In the vicinity of the points $\Delta=n\omega_m$ [including the detuning points like $\Delta=-0.9\omega_m$ which is outside of the AS regime as in Fig. \ref{figs1}(a)], chaos, especially the transient chaos between aperiodic and periodic motion with time \cite{chaos4}, is seen in the regime towards the high orbits realized under ultrahigh pump powers, which have been beyond the range of Fig. 4 in the main text. However, when the drive frequency is considerably tuned from the points $\Delta=\pm\omega_m$ to, for example, $\Delta=0.5\omega_m$ (close to a previously studied detuning range for optomechanical chaos \cite{chaos2}), one type of chaos emerges under a lower drive power. The drive amplitude for creating the chaotic motion in Fig. \ref{figs5}(b) is $E=6\times 10^5\kappa$, which can only make the system evolve to the orbit $n=0$ or $n=1$ depending on the initial condition, when the drive frequency is tuned back to $\Delta=-\omega_m$. Below this drive amplitude, with the examples of $E=5.5\times 10^5\kappa$ and $5.8\times 10^5\kappa$ in the lower panel of Fig. \ref{figs5}(b), one will see the fractional sidebands---$(1/4)\Omega_m$, $(1/2)\Omega_m$, $(3/4)\Omega_m$, etc---in these stabilized oscillations, if a spectrum analysis is performed. It indicates a route towards chaos along the continuously increased drive power. Moreover, the stabilized motion will become periodic again when the system is driven at a still higher $E=7\times 10^5\kappa$. At this detuning point $\Delta=0.5\omega_m$, there do not exist the regularly distributed orbits locked to the proper positions, those in the right part of Fig. 4 of the main text. Such an incompatibility of chaos and locked orbits may serve as a guide to study optomechanical dynamics. More properties of optomechanical chaos, which are beyond the scope of the current work, await to be explored further. 

\section*{Appendix V. A short note on the research process}

This research started from a student research project for the first author. In that project three mutually coupled cavities, one of which carries a mechanical breathing mode, are used to study the possibly triggered phonon lasing in the system. Moreover, there is an optical gain in another cavity, being close to the recently interested $\mathcal{PT}$-symmetric phonon laser. To her surprise, the first author found that the system's response (the finally stabilized cavity field intensity) became suddenly delayed for much longer time whenever the drive amplitude $E$ was only slighted modified in somewhere of the parameter space. To unravel this puzzle, we removed the optical gain and reduced the number of coupled cavities to two and until one, but a sudden delay of system response was still found to exist. After the reduction to this level, we perceived that it was due to a fundamental mechanism of cavity optomechanical systems, so we decided to search in the parameter space and saw the AS processes. Then, at the suggestion by the last author, we carefully examined the correlation between the field sidebands and mechanical frequency shift. It provides a further understanding of the AS scenario and the importance of the frequency shift $\delta$, which stabilizes the optomechanical oscillations for an ideal system with $\gamma_m=0$. Now, it is clear that the numerically observed sudden delay of system response is something occurring on the right side of Fig. 3(a) in the main text, where the different data points appear to be on the same vertical lines due to the critical slowing-down. To explore the possible existence of other dynamical behaviors, we continued to calculate at several sample points of $\Delta=n\omega_m$ and found the locked orbits like those in the right part of Fig. 4, and the chaos existing under lower pump powers was seen at the detuning points without such locked mechanical orbits.

\end{document}